\documentclass[english,useAMS, usenatbib]{mn2e}
\usepackage[T1]{fontenc}
\usepackage[latin1]{inputenc}
\usepackage{graphicx}
\usepackage{color}

\makeatletter

\providecommand{\tabularnewline}{\\}

\usepackage{times}
\usepackage{amssymb}

\newcommand{\aap}{A\&A}
\newcommand{\aaps}{A\&AS}
\newcommand{\apj}{ApJ}

\newcommand{\pasp}{PASP}
\newcommand{\aj}{AJ}
\newcommand{\mnras}{MNRAS}

\newcommand{\nat}{Nat}

\newcommand{\kmps}{\mathrm{km~s^{-1}}}
\newcommand\ion[2]{#1$\,${\sc {#2}}}   

\newcommand{\Msun}{\mathrm{M_{\odot}}}

\usepackage{babel}
\makeatother
\begin{document}
\title[Young BDs and VLM stars]{Radial and rotational velocities of young brown dwarfs and very low-mass stars in the Upper Scorpius OB association and the $\rho$~Ophiuchi cloud core}

\author[R. Kurosawa et\,al.]{Ryuichi
  Kurosawa$^1$\thanks{E-mail:rk@astro.ex.ac.uk}, Tim J. Harries$^1$
  and S.~P.~Littlefair$^2$\\ $^1$School of Physics, University of
  Exeter, Stocker Road, Exeter EX4~4QL. \\ $^2$Department of Physics
  and Astronomy, University of Sheffield, Sheffield S3 7RH}

\date{Dates to be inserted}

\pagerange{\pageref{firstpage}--\pageref{lastpage}} \pubyear{2006}

\maketitle

\label{firstpage}

\begin{abstract}

We present the results of a radial velocity (RV) survey of 14 brown
dwarfs (BDs) and very low-mass (VLM) stars in the Upper Scorpius OB
association (UScoOB) and 3 BD candidates in the $\rho$~Ophiuchi dark
cloud core. We obtained high-resolution echelle spectra at the
Very Large Telescope using Ultraviolet and Visual Echelle
Spectrograph (UVES) at two different epochs for each object,
and measured the shifts in their RVs to
identify candidates for binary/multiple systems in the sample. The
average time separation of the RV measurements is 21.6~d, and our
survey is sensitive to the binaries with separation $<0.1$~au.  We
found that  4 out of 17 objects (or $24^{+16}_{-13}$~per~cent by fraction)
show a significant RV change in 4--33~d time scale, and are 
considered as binary/multiple `candidates.'  We found no double-lined
spectroscopic binaries in our sample, based on the shape of
cross-correlation curves. The RV dispersion of the objects in UScoOB
is found to be very similar to that of the BD and VLM stars in
Chamaeleon~I (Cha~I). We also found the distribution of the mean
rotational velocities ($v\,\sin i$) of the 
UScoOB objects is similar to that of the Cha~I, but the dispersion of
$v\,\sin i$ is much larger than that of the Cha~I objects. 

\end{abstract}


\section{Introduction }

\label{sec:Introduction}

Most stars are member of binary systems and it is therefore important that
a complete star formation theory be able to predict the binary fraction,
period distribution, and mass-ratio distribution of  stellar
objects across a wide range of masses. Furthermore, the study of individual
binary systems is the only direct means to determine fundamental
stellar properties such as stellar masses and radii. 

Recent high-resolution imaging studies of brown dwarfs (BDs)
and very low-mass (VLM) stars have placed strong constraints on binaries
with separations of $\sim1-100\,\mathrm{au}$. For example, Hubble
Space Telescope (\emph{HST}) observations of $\alpha$~Per and the
Pleiades indicates a binary fraction ($f$) of $>10$~per~cent with
a bias towards separations ($a$) of less than $15$~au, and a mass-ratio
($q$) of $>0.7$ \citep{martin:2003} for objects around and below
the hydrogen burning limit (see also \citealt{bouy:2006}). A similar
lack of wide binaries was found in the field T-dwarf study \citep{burgasser:2003},
while $f\approx15$~per~cent (for the field objects with
0.03--0.1~$\mathrm{M_{\sun}}$ and M8.0--L0.5) was  determined by
\citet{close:2003} using   
the adaptive optics at Gemini North. They also found the vast majority
of systems have a semimajor axis $<20\,\mathrm{au}$ (see also
\citealt{siegler:2005} but note too the results of \citealt{luhman:2004}). An
\emph{HST} study of more than 80 field late M and L dwarfs \citep{gizis:2003}
indicated $f\approx15$~per~cent with separations in the range of 1.6--16~au.
For a small (12) sample of BDs and VLM stars (0.04--0.1~$\mathrm{M_{\odot}}$)
in Upper Scorpius OB association (UScoOB), \citet*{kraus:2005} found
$f=25_{-8}^{+16}$~per~cent for $5\,\mathrm{au}<a<18\,\mathrm{au}$
by using a similar imaging technique. More recently,
\citet{basri:2006}, combined with the results of earlier works, found
the upper limit of the overall binary fraction for VLM stars of
$26\pm10$~per~cent.   

Using a Monte Carlo simulation, the data from radial velocity
surveys available in the literature, and by carefully considering the sensitivity
and sampling biases, \citet{maxted:2005}  found an overall BD/VLM
binary frequency of 32--45 per~cent assuming $f=15$~per~cent for
$a>2.6$~au. A recent photometric study \citep{pinfield:2003} of
low-mass objects in Pleiades and Praesepe suggested, albeit indirectly,
$f$ as large as 50~per~cent. which would only be compatible with
direct imaging studies if 70--80~per~cent of those binaries have
$a<1\,\mathrm{au}$. For a more comprehensive review of the current
status of BD/VLM binary fraction and the separation distribution,
readers are refer to a recent review of multiplicity studies
by \cite{burgasser:2006}.

The extensive imaging surveys provide excellent observational constraints
on wider BD+BD binaries, but it is now necessary to search for shorter
period BD+BD binaries systematically. Binaries with the separation
of less than 1~au are not resolved by current imaging techniques,
but will be detectable as spectroscopic binaries, providing the mass-ratio
is not too extreme, and velocity separation is large enough. The first
BD+BD spectroscopic binary, PPl~15  \citep{basri:1999}, showed a
double-peaked cross-correlation function with a maximum velocity separation
of $>70\,\kmps$. The binary was found to have an eccentric orbit
($e=0.4$) with a period of $\sim5.8\,\mathrm{d}$. \citet{basri:1999}
suggested that the formation process of substellar objects is biased
towards smaller separation binaries based on the short period of PPl~15
and the lack of Pleiades BD binaries with separations $>40\,\mathrm{au}$.
Note that the median separation of binaries with solar-type primaries
is 30~au \citep{duquennoy:1991}. Pioneering work on the RVs
of BDs and VLMs are presented by \citet{guenther:2003}, \cite{kenyon:2005}
and \citet{joergens:2006} who found a several binary candidates;
however, the orbital parameters and masses of binaries remains
unknown because the follow-up spectroscopic monitoring is lacking
or is still being undertaken. In addition to the follow-up observations,
the number of BDs and VLM star binary candidates needs to be increased
in order to have better statistics on short-period binary parameters. 

The first BD+BD eclipsing binary (2MASS J0532184--J0546085) was
discovered by \citet*{stassun:2006} from the $I$-band photometric
monitoring of the system. Combining their light curves and the results
from the follow-up radial velocity measurements, they were able to determined
the precise orbital and physical parameters of the system.  The projected
semimajor axis and the period of the binary are found as
$0.0398\pm0.0010$~au and $9.779621\pm0.000042$~d respectively. 

The separation distribution of BD/VLM binaries is critical to
understanding their origin. There are two main models for the
formation of BDs and VLM stars: first, they have low masses because
they form in low-mass, dense molecular cloud cores
(e.g.~\citealt{padoan:2002}); second, BD/VLM objects have low masses
because they are ejected from the dense core in which they form via
dynamical interactions in multiple system, cutting off their accretion
before they have reached stellar masses (\citealt{reipurth:2001};
\citealt*{bate:2002}). Alternatively, there is a third model in which
a free-floating BD or planetary-mass object can be formed in the
process of the photo-evaporation
(e.g.~\citealt{mccaughrean:2002};~\citealt{whitworth:2004}) with  
the outer layers of a pre-stellar core ($\sim0.2\,\mathrm{M_{\odot}}$)
removed by the strong radiation pressure from the nearby massive OB
stars before the accretion onto the protostar at core centre occurs.

Due to the dynamical interaction involved in the second model, BD/VLM
binaries that survive are generally expected to have small separations.
In the first model, wider binaries may be expected to be more common.
\citet*{bate:2002b} suggested that close binaries ($a<10\,\mathrm{au}$)
do not form directly, but result from hardening of wider systems though
a combination of dynamical interactions, accretion and interactions
with circum-binary discs. If BD/VLM binaries have formed through such
mechanisms, one would not expect to find binaries with 1--10~au separations
without also finding many with separation $<1$~au. If an absence/rarity
of binaries with 1~au were found, it may support the idea that
they are ejected quickly from multiple systems before they have undergone
the interactions that shorten their periods. 

Our immediate aim is to identify spectroscopic and close BD/VLM binaries
using the high-resolution echelle spectroscopy at two epochs. This
experiment is sensitive to VLM binaries with separations of $<0.1$~au
which corresponds to a period of $\sim10\,\mathrm{d}$. A larger sample
of candidates will enable us to measure the binary fraction of these
short-period/close binaries (once confirmed), and address whether
there is a significant population of `hidden' VLM companions. The
long term goal of this project to follow up the binary candidates
found in this paper by spectroscopically monitoring them over different
time scales, enabling us to obtain the radial velocity curves and
their minimum masses. 

In Section~\ref{sec:observation-reduction}, we describe the
observations and the data reduction. The results of radial velocity
and rotational velocity ($v\,\sin i$) measurements are presented in
Section~\ref{sec:Results}. We discuss the binary/multiplicity fraction
indicated by our RV survey in Section~\ref{sec:Binary-fraction}, and
give our conclusions in Section~\ref{sec:Conclusions}.

\begin{table*}

\caption{Summary of known properties of the targets from literature: a.~\citet{luhman:1999}
(original list for $\rho$~Oph), b.~\citet*{ardila:2000} (original
list for UScoOB), c.~\citet*{wilking:1999}, d.~\citet{muzerolle:2003},
e.~\citet{kraus:2005}, and f.~\citet*{mohanty:2005}. }

\label{tab:literatures}

\begin{center}\begin{tabular}{llcccc}
\hline 
Object&
Sp.&
mass&
RV&
$v\sin i$&
Known multiple?\tabularnewline
&
&
$\left[\mathrm{M_{\odot}}\right]$&
$\left[\mathrm{km\, s^{-1}}\right]$&
$\left[\mathrm{km\, s^{-1}}\right]$&
\tabularnewline
\hline 
GY 5&
$\mathrm{M7^{c}}$&
$0.07^{\mathrm{d}}$&
$-6.3\pm1.9^{\mathrm{d}}$&
$16.8\pm2.7^{\mathrm{d}}$&
no\tabularnewline
GY 141&
$\mathrm{M8.5^{a}}$&
$0.02^{\mathrm{d}}$&
\ldots{}&
$6.0^{f}$&
no\tabularnewline
GY 310&
$\mathrm{M8.5^{c}}$&
$0.08^{\mathrm{a,d}}$&
\ldots{}&
$10.0^{f}$&
no\tabularnewline
USco 40&
$\mathrm{M5^{b}}$&
$0.1^{\mathrm{b}}$&
\ldots{}&
$37.5^{f}$&
no\tabularnewline
USco 53&
$\mathrm{M5^{b}}$&
$0.1^{\mathrm{b}}$&
\ldots{}&
$45.0^{f}$&
no\tabularnewline
USco 55&
$\mathrm{M5.5^{b}}$&
$0.10+0.07^{\mathrm{e}}$&
\ldots{}&
$12.0^{f}$&
$\mathrm{visual^{e}}$\tabularnewline
USco 66&
$\mathrm{M6^{b}}$&
$0.07+0.07^{\mathrm{e}}$&
$-4.4\pm0.6^{\mathrm{d}}$&
$27.5^{f}$&
$\mathrm{visual^{e}}$\tabularnewline
USco 67&
$\mathrm{M5.5^{b}}$&
$0.10^{\mathrm{e}}$&
\ldots{}&
$18.0^{f}$&
no\tabularnewline
USco 75&
$\mathrm{M6^{b}}$&
$0.07^{\mathrm{e}}$&
$-5.6\pm1.1^{\mathrm{d}}$&
$63.0^{f}$&
no\tabularnewline
USco 100&
$\mathrm{M7^{b}}$&
$0.05^{\mathrm{e}}$&
$-8.9\pm0.6^{\mathrm{d}}$&
$50.0^{f}$&
no\tabularnewline
USco 101&
$\mathrm{M5^{b}}$&
$0.05^{\mathrm{b}}$&
\ldots{}&
\ldots{}&
no\tabularnewline
USco 104&
$\mathrm{M5^{b}}$&
$0.05^{\mathrm{b}}$&
\ldots{}&
$16.0^{f}$&
no\tabularnewline
USco 109&
$\mathrm{M6^{b}}$&
$0.07+0.04^{\mathrm{e}}$&
$-3.8\pm0.7^{\mathrm{d}}$&
$6.0^{f}$&
$\mathrm{visual^{e}}$\tabularnewline
USco 112&
$\mathrm{M5.5^{b}}$&
$0.1^{\mathrm{e}}$&
\ldots{}&
$8.0^{f}$&
no\tabularnewline
USco 121&
$\mathrm{M6^{b}}$&
$0.02^{\mathrm{b}}$&
$-38.9\pm1.0^{\mathrm{d}}$&
\ldots{}&
no\tabularnewline
USco 128&
$\mathrm{M7^{b}}$&
$0.05^{\mathrm{e}}$&
$-3.0\pm1.6^{\mathrm{d}}$&
$0.0^{f}$&
no\tabularnewline
USco 130&
$\mathrm{M7.5^{e}}$&
$0.04^{\mathrm{e}}$&
\ldots{}&
$14.0^{f}$&
no\tabularnewline
USco 132&
$\mathrm{M7^{b}}$&
$0.05^{\mathrm{e}}$&
$-8.2\pm1.1^{\mathrm{d}}$&
\ldots{}&
no\tabularnewline
\hline
\end{tabular}\par\end{center}

\end{table*}


\section{Observations}

\label{sec:observation-reduction}

Our sample consists of 18 young, very low-mass objects: 15 objects in
the Upper Scorpius OB association ($\mathrm{d\approx145\, pc}$,
\citealt{dezeeuw:1999}) from the list of \citet*{ardila:2000} and 3
objects in the $\rho$~Ophiuchi cloud core ($\mathrm{d\approx150\,
pc}$, \citealt{dezeeuw:1997}) from \citet{luhman:1999}. The spectral
type of the objects range between M5 and M8.5, and the age
$<\sim10\,\mathrm{Myr}$ (\citealt{luhman:1999}; \citealt{ardila:2000};
\citealt{muzerolle:2003}; \citealt{kraus:2005}).  The sample is not
complete, and the selection was solely based on brightness and the
observability. The basic properties of the targets based on the
literature is summarised in Table~\ref{tab:literatures}.

We obtained high-resolution spectra with the Kueyen telescope of the Very Large Telescope
(VLT--Cerro Parnal, Chile) using the UVES echelle spectrograph. The
observations were carried out between 2004 April 5 and 2004 May 17 in
 service mode. For each object, spectra were obtained at two different
epochs separated by 4--33~d. For each object on a
given night, two separate spectra were obtained consecutively. This
allows us to derive more reliable uncertainty estimates in the RV
values of our targets (c.f.~\citealt{joergens:2006}).  The data were
obtained using the red arm of UVES spectrograph with two mosaic CCDs
(EEV + MIT/LL with 2k$\times$4k pixels). The wavelength coverage of
6708 -- 10,250~\AA\, and the spectral resolution $R\approx40,000$ were
used. The slit width and length of 1'' and 12'' were used respectively
with a typical seeing of 0.8''.

The data were reduced via the standard ESO pipeline procedures for
UVES echelle spectra. In summary, the data were corrected for bias,
interorder background, sky background, sky emission lines and cosmic
ray hits. They were then flattened, optimally extracted, and finally
the different orders were merged. No binning was performed to achieve
high resolution required for the RV measurements. The wavelength was
calibrated using the Thorium-Argon arc spectra with a typical value
of the standard deviation of the dispersion solution of 5~m\AA\,
which corresponds to 0.2~$\kmps$ at the central wavelength 8600~\AA.
However, the autoguiding of the telescope keeps the star at the centre
of the slit with about a tenth of the FWHM ($1\,\kmps$) which sets
the upper limit for the systematic error in the RV measurements
\citep{bailer-jones:2004}.
\textcolor{black}{
In fact, we find that any systematic error is negligible compared to
the random error associated with our radial velocity measurements (see
Section~\ref{sub:Radial-velocities}).
}
A typical
signal-to-noise ratio (S/N) per wavelength bin of the spectra is about
15, and the heliocentric velocity correction was applied to the final
spectra.

\textcolor{black}{
In addition to the main targets in Table~\ref{tab:literatures}, we
also obtained the spectra of LHS~49 (on 2004-April-17) and HD~140538
(on 2004-May-06), which are the RV template and the RV standard
stars respectively.  These data were obtained using the same telescope
and the same instrument setup as for our main targets.  The S/N of
the both objects were about 60. 
}

\section{Results}

\label{sec:Results}

\subsection{Radial velocities}

\label{sub:Radial-velocities}

\begin{table*}

\caption{Summary of the observations, the heliocentric radial velocities ($\mathrm{RV}$)
from two-epoch and the average rotational velocities ($v\sin i$).
The uncertainties of relative radial velocities ($\sigma_{\mathrm{RRV}}$)
with respect to the template star LHS~049 and the average radial
velocities ($\overline{\mathrm{RV}}$) are also given.  
\textcolor{black}{
N.B.~the uncertainties listed along with the heliocentric radial
velocities in column~4 and 6 include the uncertainty in the
heliocentric RV of the template star, but $\sigma_{\mathrm{RRV}}$ in
column~5 do not.
}
}

\label{tab:summary_results}

\begin{center}\begin{tabular}{lcrrrrr}
\hline 
Object&
Date&
HJD-2453100&
RV&
$\sigma_{\mathrm{RRV}}$&
$\overline{\mathrm{RV}}$&
$v\sin i$\tabularnewline
&
&
&
$\left[\mathrm{km\, s^{-1}}\right]$&
$\left[\mathrm{km\, s^{-1}}\right]$&
$\left[\mathrm{km\, s^{-1}}\right]$&
$\left[\mathrm{km\, s^{-1}}\right]$\tabularnewline
\hline 
GY 5&
2004-Apr-24&
20.7198440&
$-6.77\pm2.02$&
1.96&
&
\tabularnewline
&
2004-May-07&
33.7721938&
$-6.02\pm0.85$&
0.68&
$-6.39\pm2.19$&
$16.5\pm0.6$\tabularnewline
GY 141&
2004-May-10&
36.6712108&
$-4.15\pm0.63$&
0.38&
&
\tabularnewline
&
2004-May-17&
43.6683007&
$-3.46\pm0.66$&
0.43&
$-3.81\pm0.91$&
$4.4\pm1.4$\tabularnewline
GY 310&
2004-Apr-24&
20.8373504&
$-5.27\pm0.92$&
0.77&
&
\tabularnewline
&
2004-May-09&
35.8224052&
$-8.19\pm0.71$&
0.51&
$-6.73\pm1.16$&
$11.1\pm6.0$\tabularnewline
USco 40&
2004-Apr-05&
1.7661754&
$-7.34\pm0.52$&
0.15&
&
\tabularnewline
&
2004-May-07&
33.7620554&
$-6.28\pm0.52$&
0.15&
$-6.81\pm0.74$&
$34.2\pm0.5$\tabularnewline
USco 53&
2004-Apr-04&
0.9036653&
$-8.27\pm0.70$&
0.50&
&
\tabularnewline
&
2004-May-02&
28.7394676&
$-7.22\pm0.63$&
0.39&
$-7.75\pm0.95$&
$40.0\pm0.6$\tabularnewline
USco 55&
2004-Apr-05&
1.8422807&
$-6.21\pm0.50$&
0.04&
&
\tabularnewline
&
2004-May-02&
28.8141198&
$-6.55\pm0.50$&
0.06&
$-6.38\pm0.71$&
$22.9\pm0.8$\tabularnewline
USco 66&
2004-Apr-05&
1.7972634&
$-7.24\pm0.95$&
0.81&
&
\tabularnewline
&
2004-May-02&
28.7956003&
$-8.38\pm1.32$&
1.22&
$-7.81\pm1.63$&
$25.9\pm1.2$\tabularnewline
USco 67&
2004-Apr-05&
1.7188620&
$-7.07\pm0.67$&
0.44&
&
\tabularnewline
&
2004-May-02&
28.7113799&
$-5.70\pm0.58$&
0.30&
$-6.38\pm0.88$&
$18.4\pm0.4$\tabularnewline
USco 75&
2004-Apr-04&
0.8840376&
$-6.80\pm0.70$&
0.50&
&
\tabularnewline
&
2004-May-07&
33.6065432&
$-6.16\pm0.78$&
0.60&
$-6.48\pm1.05$&
$55.6\pm3.0$\tabularnewline
USco 100&
2004-Apr-06&
1.8179138&
$-7.37\pm0.67$&
0.44&
&
\tabularnewline
&
2004-May-02&
28.7752928&
$-7.47\pm0.62$&
0.36&
$-7.42\pm0.91$&
$43.7\pm3.2$\tabularnewline
USco 101&
2004-Apr-04&
0.8120734&
$-3.25\pm0.66$&
0.44&
&
\tabularnewline
&
2004-May-02&
28.6591660&
$-5.03\pm0.60$&
0.32&
$-4.14\pm0.89$&
$19.1\pm0.3$\tabularnewline
USco 104&
2004-Apr-04&
0.7850385&
$-7.05\pm0.61$&
0.35&
&
\tabularnewline
&
2004-May-02&
28.6349480&
$-7.62\pm0.57$&
0.27&
$-7.33\pm0.83$&
$16.7\pm0.4$\tabularnewline
USco 109&
2004-Apr-05&
1.7453989&
$-4.72\pm0.50$&
0.07&
&
\tabularnewline
&
2004-May-07&
33.6304878&
$-4.70\pm0.51$&
0.09&
$-4.71\pm0.72$&
$8.6\pm1.2$\tabularnewline
USco 112&
2004-Apr-04&
0.8552168&
$-2.95\pm0.50$&
0.05&
&
\tabularnewline
&
2004-May-07&
33.5826368&
$-3.41\pm0.51$&
0.07&
$-3.18\pm0.71$&
$5.8\pm1.2$\tabularnewline
USco 121&
2004-Apr-24&
20.7059408&
$-40.76\pm0.90$&
0.74&
&
\tabularnewline
&
2004-May-02&
28.6969641&
$-42.48\pm0.54$&
0.20&
$-41.62\pm1.04$&
$17.6\pm1.3$\tabularnewline
USco 128&
2004-May-13&
39.7978276&
$-7.03\pm0.51$&
0.09&
&
\tabularnewline
&
2004-May-17&
43.6108024&
$-6.32\pm0.51$&
0.10&
$-6.68\pm0.72$&
$3.6\pm1.1$\tabularnewline
USco 130&
2004-May-09&
35.7724090&
$-3.92\pm0.94$&
0.80&
&
\tabularnewline
&
2004-May-13&
39.8538830&
$-3.90\pm0.67$&
0.45&
$-3.91\pm1.16$&
$15.2\pm1.1$\tabularnewline
USco 132&
2004-May-13&
39.8268683&
$-6.68\pm0.53$&
0.18&
&
\tabularnewline
&
2004-May-17&
43.6391138&
$-6.61\pm0.54$&
0.20&
$-6.64\pm0.76$&
$9.1\pm0.7$\tabularnewline
\hline
\end{tabular}\par\end{center}

\end{table*}



\begin{figure*}

\begin{center}

\includegraphics[%
  clip,
  scale=0.9]{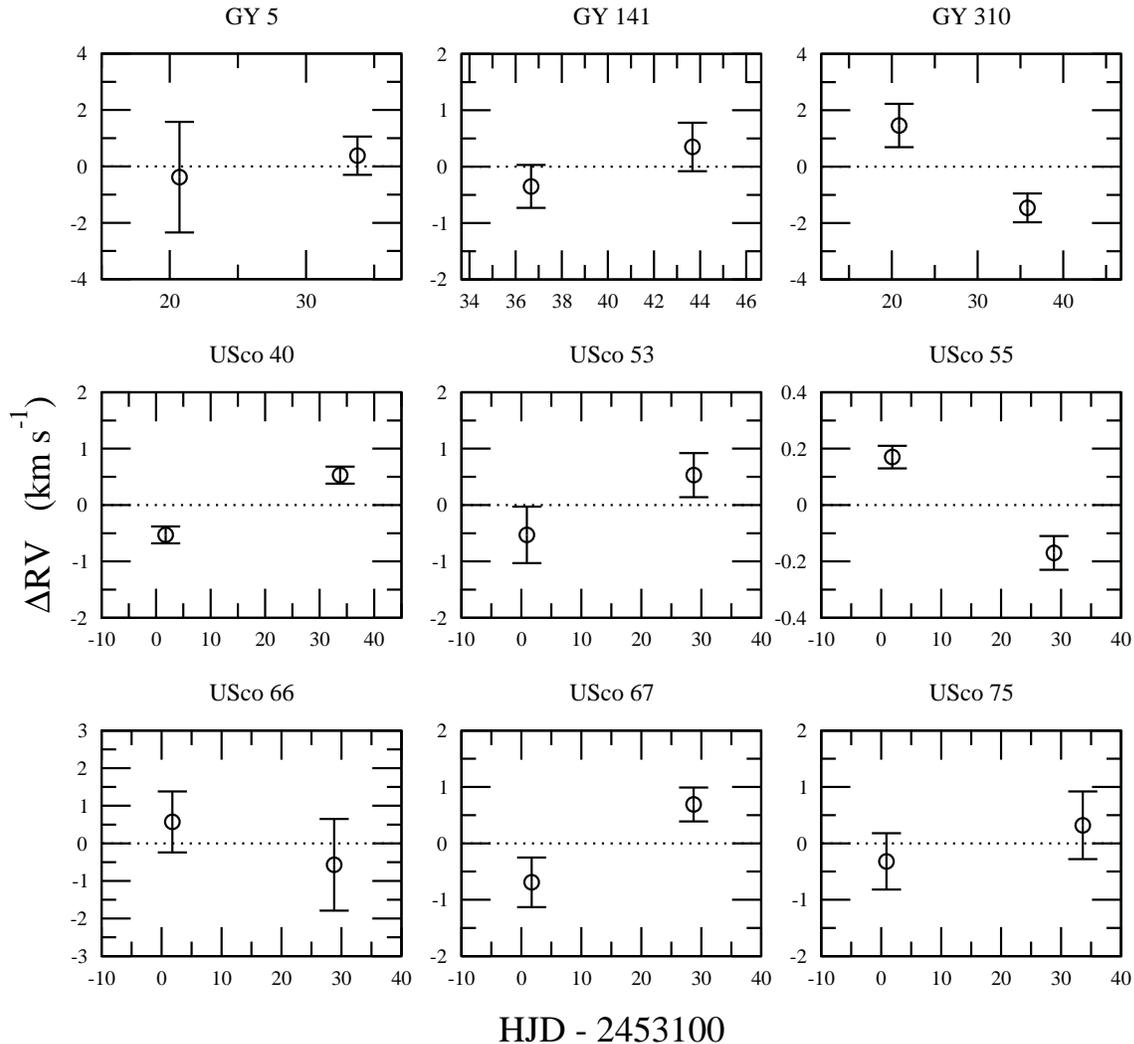}

\end{center}

\caption{Relative radial velocities (RVs) of objects measured in two
different epochs. The vertical axes indicate the amount of deviation
($\Delta \mathrm{RV}$) from the `average' radial velocity
($\overline{\mathrm{RV}}$) in Table~\ref{tab:summary_results}, and
the horizontal axes indicate the time of the observation in
heliocentric Julian date (HJD). The objects are considered to have a
non-constant RV when the error bars of two data points do not overlap
each other. The non-constant RV objects are considered as
binary/multiple candidates.}

\label{fig:radvel01}

\end{figure*}


\addtocounter{figure}{-1}


\begin{figure*}

\begin{center}

\includegraphics[%
  clip,
  scale=0.9]{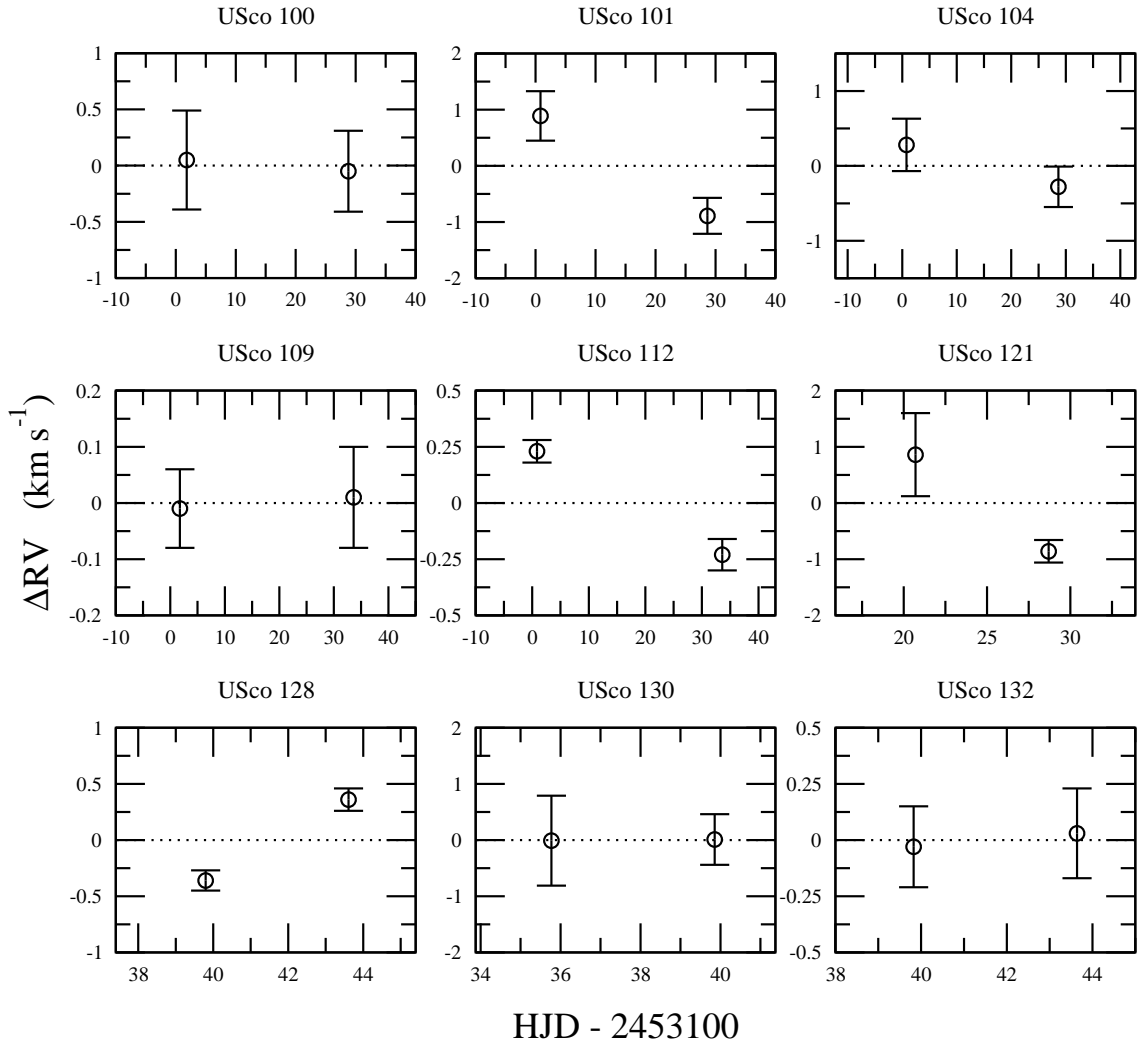}

\end{center}

\caption{continued}

\label{fig:radvel02}

\end{figure*}


The radial velocities of each object were determined by using the cross-correlation
function of the object spectrum with that of a template star which has
a similar spectra type. By visual inspection,
the wavelength ranges used for the cross-correlation calculations
are chosen by avoiding the regions of spectra affected by the telluric
lines, and defects and fringes (in near infrared) of the CCDs. The
radial velocities of objects with respect to the template are obtained
by measuring the location of the peak in the cross-correlation
function. The location of the peak is determined by fitting the
section of the cross-correlation function around the peak by a second
order polynomial. LHS~49 (Proxima Cen, M5.5) was
chosen as the template for this purpose. The heliocentric radial velocity of the
template object LHS~49 was obtained by measuring the wavelength shifts
of the prominent photospheric absorption features \ion{K}{i}~$\lambda\lambda$7664.911,~7698.974.
This gives us $\mathrm{RV_{LHS49}=}-22.6\pm0.5\,\kmps$, which is
in good agreement with the earlier measurement of \citet{garcia-sanchez:2001}
who found $\mathrm{RV_{LHS49}=}-21.7\pm1.8\,\kmps$. The heliocentric
RV of each object can be then calculated by adding $\mathrm{RV_{LHS49}}$
with the RV of each object with respect to LHS~49. In the following
measurements of the heliocentric radial velocities, our measurement
($\mathrm{RV_{LHS49}=}-22.6\pm0.5\,\kmps$) will be used for consistency. 

Before applying the cross-correlation technique to our main targets,
we have applied the technique to the radial velocity standard
HD~140538 \textcolor{black}{(G2.5V)}
for which an high-accuracy RV measurement via the fixed-configuration,
cross-dispersed \'{e}celle spectrograph Elodie \citep{baranne:1996}
is available. This was done so to ensure not only the validity of
the cross-correlation technique, but also the validity of the wavelength
calibration. In this test, we found the heliocentric 
\textcolor{black}{
$\mathrm{RV_{HD\,140538}=}19.17\pm0.40\,\kmps$
}
which is in good agreement with the Elodie radial velocity measurement
of $19.00\pm0.05\,\mathrm{km\, s^{-1}}$ \citep*{udry:1999}.
\textcolor{black}{
  Note that the uncertainty in $\mathrm{RV_{HD\,140538}}$ was computed
  by combining (in quadrature) the uncertainty ($0.4~\kmps$) in the
  heliocentric RV of the template star HD~74497 (G3V) and the
  uncertainty in the relative RV of HD~140538 ($0.02~\kmps$), measured
  by cross correlation with the template star.  Clearly, the former
  dominates in the uncertainty of
  $\mathrm{RV_{HD\,140538}}$ shown above. The uncertainty in the
  heliocentric RV of the template is rather large because it was
  determined from a few photospheric lines.
}

\textcolor{black}{
  The results of the heliocentric RV measurements (from two epochs for
  each object) are summarised in Table 2 along with their
  uncertainties.  The uncertainties listed in columns~4 and 6 include
  the uncertainty of the absolute RV of the template star, added in
  quadrature with the uncertainty found from the cross correlation
  analysis. On the other hand, the uncertainties in the relative
  radial velocities ($\sigma_{\mathrm{RRV}}$) listed in column~5 of
  the table \emph{do not} include the uncertainty of the template star. It is
  $\sigma_{\mathrm{RRV}}$ that is relevant for the identification of
  RV variable objects.
}
 
To determine the $\sigma_{\mathrm{RRV}}$, we have used the following
procedure: (1)~add random gaussian deviate noise to the object spectra using the
corresponding variance spectra\footnote{\textcolor{black}{Defined by the variance of the
$\chi^{2}$ fit to the signal obtained during optimum extraction \citep[c.f.][]{horne:1986}.}}, 
(2)~compute the cross-correlation
curve using the spectra with added noise and the template, (3)~measure
the RV by locating the peak, (4)~repeat 1--3 for 100 times, and compute the standard
deviation of the 100 RV values. This should provide a good estimate of
uncertainties in RVs cased by the uncertainties in the flux levels.
We also estimated the RV uncertainties using the standard deviations on the
mean of the two independent RV measurements obtained from the two
consecutive spectra taken on the same night (c.f.~\citealt{joergens:2006}).
We found that these estimates agreed well with those obtained from
the Monte Carlo method. The `average' radial velocities $\left (\overline{\mathrm{RV}} \right )$ of the
two epochs are also given in Table~\ref{tab:summary_results}. 

For each object and for each RV measurement, the deviations ($\Delta\mathrm{RV}$)
from the average RV are computed and summarized in Fig.~\ref{fig:radvel01}
along with their uncertainties ($\sigma_{\mathrm{RRV}}$) in order
to aid the identification of multiplicity candidates. Note that in
computing $\Delta\mathrm{RV}$ we do not require the knowledge of
absolute\textcolor{black}{/heliocentric} radial velocities, but only the relative
velocities (with respect to a template). 
We also  cross-correlated object spectra from the two epochs 
(instead of using the spectra of the template LHS~49)  in order to find
$\Delta\mathrm{RV}$. The resulting cross-correlation functions were
found to be too noisy (due to the relatively low S/N in the object
spectra) for the  peak positions of the cross-correlation curves to be
reliably located. 

To identify an object with a RV variation with a statistical
significance from our sample, we apply the method described by
\citet{maxted:2005}, which we briefly summarize next, to our data.
There are three steps in this method: 
(1)~compute the $\chi^{2}$ by fitting the two-epoch RV data for each 
object with a constant function (a zero-th order polynomial),
(2)~compute the corresponding $\chi^{2}$ probability ($p$),
(3)~designate the object as a non-constant RV object or a binary
candidate if $p<10^{-3}$ (\textcolor{black}{0.1~per~cent}).  When computing $\chi^{2}$, we
use the uncertainties in the relative RV ($\sigma_{\mathrm{RRV}}$). 
The number of degree of freedom in the fitting procedure is obviously
1.  A similar method was also used in a recent RV survey of VLM stars
by \citet{basri:2006}. 

\begin{figure}

\begin{center}
\includegraphics[%
  clip,
  scale=0.46]{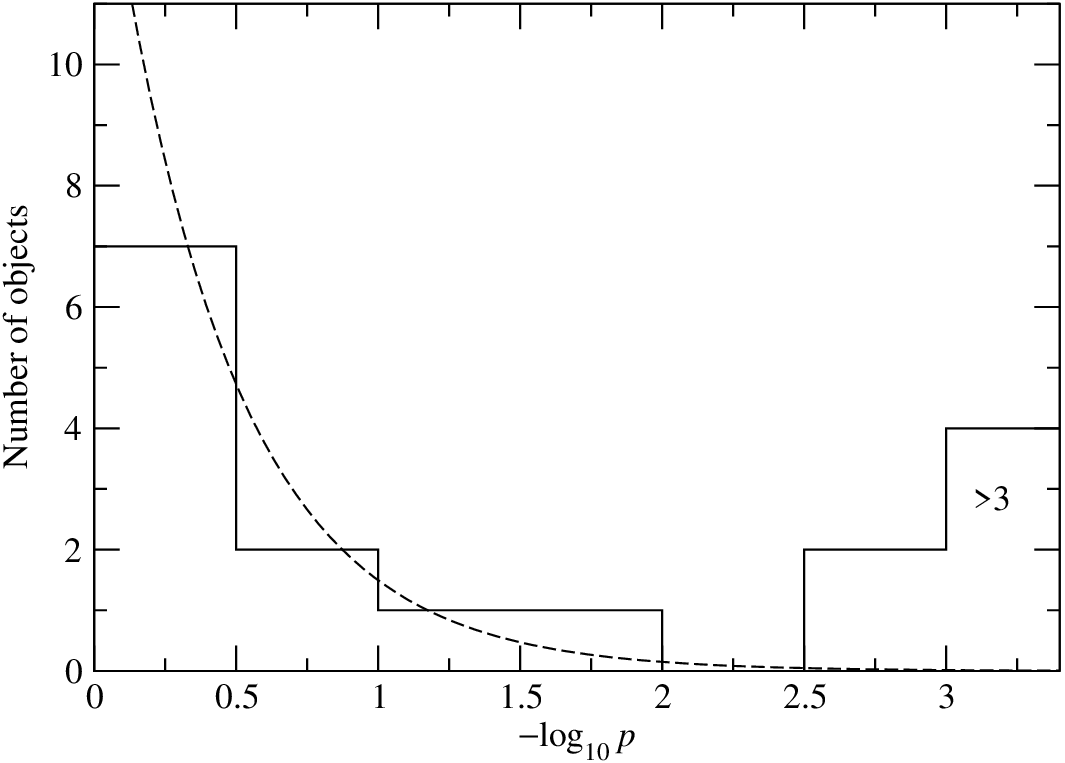}
\end{center}

\caption{Histogram of the $\chi^2$ probability ($p$) for fitting the
  observed (relative) RV values with a constant (a horizontal
  line). The constant used in the fit is determined from the weighted mean of
  the two RV measurements for each object.  The objects with
  $p<10^{-3}$ are identified as \emph{non-constant} or \emph{multiple},
  which appear in the right-most bin in the histogram. There
  are four objects which satisfy this condition. 
  \textcolor{black}{The expected distribution (dashed) for RV constant
  objects is also shown for a comparison, and is 
  normalized to the total number of objects that have $p>10^{-3}$}.
  The match between the expected distribution and the
  histogram is reasonable \textcolor{black}{(c.f.~Section~\ref{sub:Radial-velocities})},
  indicating that our uncertainty estimates 
  in RV values are also reasonable.  }

\label{fig:chisq-test}

\end{figure}

We have computed $p$ for all the objects (Table~\ref{tab:chisq_prob}),
and have plotted the results  
as a histogram of $-\log{p}$, shown in Fig.~\ref{fig:chisq-test}
(excluding the non-member USco~121; see explanation later). 
The figure clearly shows two distinctive populations: one on the left
(with small $-\log{p}$ values) occupied by the RV constant objects,
and one on the right (with the large $-\log{p}$ values)
occupied by the RV variable candidates.  The
expected distribution of $p$ computed consistently with our
uncertainty measurements is also shown in the same figure. The RV
constant population on the left side reasonably matches the expected
curve. 
\textcolor{black}{
To quantify this point, we compute the
$\chi^{2}$ probability to test the goodness of the fit of the expected
distribution to the histogram. We restrict the fit to the first 4 bins
from the left $\left(0 < -\log_{10}p < 2 \right)$ of the
histogram since the objects in the second bin from the right
$\left(2.5 < -\log_{10}p < 3 \right)$ are `potentially'
RV-variable, although we have flagged them as the RV-constant
objects. In this analysis, we find the probability of 0.5 which
indicates that the fit is reasonable. Further, if we re-normalized the
expected distribution, accounting for that fact we neglected the
objects in the second bin from the right, we find the $\chi^{2}$
probability of the fit increases to 0.9.  This, in turn, indicates
that our uncertainties in RV values are reasonable, and an additional
systematic error is not necessary in this analysis. 
}
\textcolor{black}{
This assertion is supported by the re-analysis of BD RVs, obtained by
\citet{joergens:2006} using the same instruments setup
as ours, by \citet{maxted:2005}. 
}

The boundary ($p=10^{-3}$) between the RV variable and the RV
constants seems somewhat arbitrary, but here we simply adopt the
definition of \citet{maxted:2005}.  Based on this criteria, there are
four objects which show significant RV variations (out of 17 samples)
as one can see from the figure.   They are USco~112, USco~128,
USco~40, and USco~55, and considered as our preliminary binary/multiple 
candidates.  We will discuss the binary fraction and the expected binary
detection probability later in Section~\ref{sec:Binary-fraction}.


\begin{table}

\caption{The $\chi^{2}$ probabilities ($p$) of each object being a RV constant,
listed in ascending order of $p$. The ellipses represent the boundary
($p=10^{-3}$) between RV constant objects and RV variable objects. }

\label{tab:chisq_prob}

\begin{center}\begin{tabular}{cc}
\hline 
obj.~ID&
$p$\tabularnewline
\hline 
USco~112&
$8.9\times10^{-8}$\tabularnewline
USco~128&
$1.3\times10^{-7}$\tabularnewline
USco~40&
$5.8\times10^{-7}$\tabularnewline
USco~55&
$2.4\times10^{-6}$\tabularnewline
$\cdots$&
$\cdots$\tabularnewline
USco~101&
$1.1\times10^{-3}$\tabularnewline
GY~310&
$1.6\times10^{-3}$\tabularnewline
USco~67&
$1.0\times10^{-2}$\tabularnewline
USco~53&
$9.8\times10^{-2}$\tabularnewline
USco~104&
$2.0\times10^{-1}$\tabularnewline
GY~141&
$2.3\times10^{-1}$\tabularnewline
USco~75&
$4.1\times10^{-1}$\tabularnewline
USco~66&
$4.4\times10^{-1}$\tabularnewline
GY~5&
$7.2\times10^{-1}$\tabularnewline
USco~132&
$7.9\times10^{-1}$\tabularnewline
USco~100&
$8.6\times10^{-1}$\tabularnewline
USco~109&
$8.6\times10^{-1}$\tabularnewline
USco~130&
$9.8\times10^{-1}$\tabularnewline
\hline
\end{tabular}\par\end{center}

\end{table} 

\begin{figure}

\begin{center}

\includegraphics[%
  clip,
  scale=0.46]{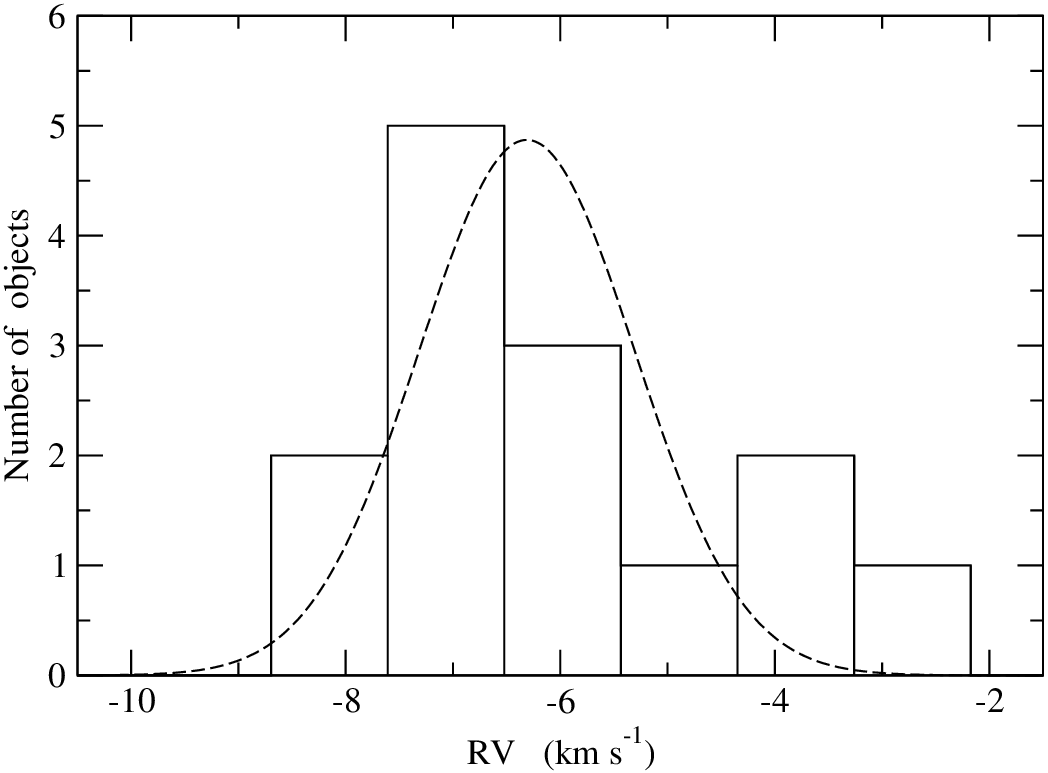}

\end{center}

\caption{Histogram of the average heliocentric radial velocities of 14 UScoOB BD and VLM
objects listed in Table~\ref{tab:summary_results} (excluding USco~121,
a non-member). The gaussian fit (dashed) of the radial velocity
distribution gives a standard deviation of $1.0\,\kmps$ and the peak
position of $-6.3\,\kmps$. }
\label{fig:vrad_histrogram}

\end{figure}

Finally, the histogram of $\overline{\mathrm{RV}}$ for the objects in UScoOB
is given in Fig.~\ref{fig:vrad_histrogram}. The total number
of the objects is 14. Note that USco~121 is excluded from the graph
since it is identified as a non-member of the UScoOB association based
on the RV value (see Table~\ref{tab:summary_results}).  \citet{muzerolle:2003}
also found it to be a likely non-member based on the radial velocity
and the low lithium abundance. The distribution of the RVs in the
figure was fitted by a gaussian function. We found that the standard
deviation and the peak position of the radial velocity distribution
are $1.0\,\kmps$ and $-6.3\,\kmps$ respectively. The former is very
similar to the standard deviation (0.9~$\kmps$) of the radial velocity
distribution of 9 BDs and VLM objects in Cha~I found by \citet{joergens:2006b}.
They also studied the radial velocity distribution of more massive
25 T~Tauri stars in Cha~I, and found the standard deviations ($1.3\,\kmps$)
is not significantly different from that of the brown dwarfs and the
very low-mass objects. Unfortunately, we do not have the radial velocity
measurements of the higher mass counter parts (T~Tauri stars) in
Upper Sco OB association. This is planned for  future investigation
since this is important for the study of the mass dependency of the kinematics
in a young stellar cluster. 

According to the hydrodynamical simulations of a low-mass star-forming
cluster of \cite{bate:2003} which yields a stellar density of
$\sim10^{3}\,\mathrm{stars\, pc^{-3}}$, the rms dispersion (1-D)
of the stars and the BDs is $1.2\,\kmps$.  Similarly for the model
with a higher stellar density ($\sim10^{4}\,\mathrm{stars\,
pc^{-3}}$), the rms dispersion is 2.5~$\kmps$
(\citealt{bate:2005}). The standard deviation of
$\overline{\mathrm{RV}}$ ($1.0\,\kmps$) found in our analysis is more
comparable the lower stellar density model.


\subsection{Rotational velocities}

\label{sub:Rotational-velocities}

The rotational velocities of the objects were determined by measuring
the widths of the cross correlation functions of the target spectra against
a template spectrum from an object which is known to have a very small
rotational velocity. The line broadening of the targets is assumed
to be dominated by rotational broadening. As in the cases for the
radial velocity measurements, LHS~49 is chosen as the template.
Using its rotational period ($P\approx83\,\mathrm{d}$, \citealt{benedict:1998})
and radius ($R_{*}\approx0.145\,\mathrm{R_{\odot}}$ from the VLTI
measurement by \citealt{segransan:2003}), the rotational velocity
of LHS~49 is estimated as $v\,\sin i=2\pi R_{*}/P\approx0.1\,\kmps$;
negligibly small. 

The width of the cross-correlation curves ($\sigma_{\mathrm{CCF}}$)
are calibrated with the rotational velocities ($v\,\sin i$) by cross
correlating the template spectra against the same template spectra
with added rotation (convolved with a given $v\,\sin i$), as
done by e.g.~\citet{tinney:1998}, \citet{mohanty:2003} and \citet{white:2003}.
A linear limb-darkening law with a solar-like parameter ($\epsilon=0.6$)
was assumed in the formulation of the rotational profile described
by \citet{gray:1992}, his Eq.~17.12. For each object, two measurements
of rotational velocities are computed from two independent spectra
obtained at different epochs. As for the RV measurements,
the mean and the standard deviation of the mean are used as the final
rotational velocity and its uncertainty. The final results are recorded
in Table~\ref{tab:summary_results}. In general, our measurements
are in good agreement with the earlier measurements of \citet{muzerolle:2003}
and \citet*{mohanty:2005}, given in Table~\ref{tab:literatures}.
For example, \citet{muzerolle:2003} found $v\,\sin i=16.8\pm2.7\,\mathrm{km\, s^{-1}}$
for GY~5 while we found $v\,\sin i=16.5\pm0.6\,\mathrm{km\, s^{-1}}$. 

\begin{figure}

\begin{center}

\includegraphics[%
  clip,
  scale=0.63]{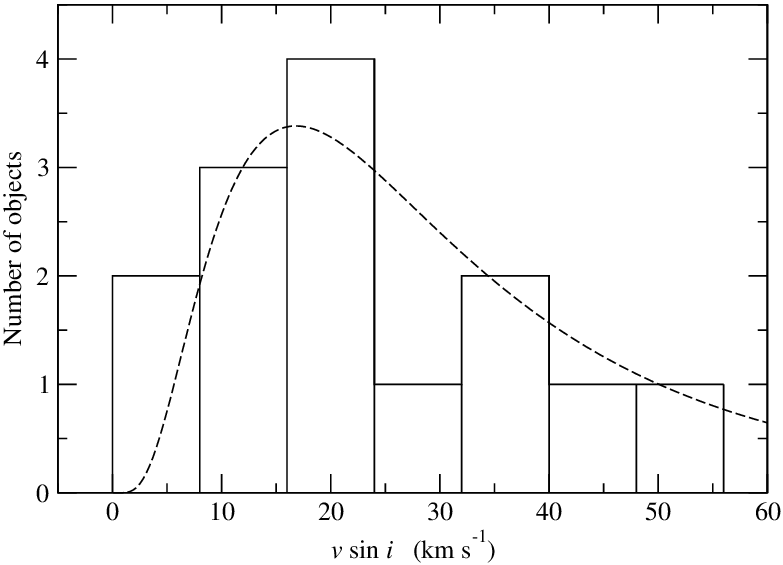}

\end{center}

\caption{Histogram of the rotational velocities ($v\,\sin{i}$) of 14 UScoOB BD and VLM
objects listed in Table~\ref{tab:summary_results} (excluding USco~121,
a non-member).  The log-normal fit (dashed) of the
rotational velocity distribution gives a standard deviation
$27.8\,\kmps$ and the peak position $16.9\,\kmps$.}
\label{fig:vrot_histrogram}

\end{figure}

The range of $v\,\sin i$ found among our objects is $3.6$--$55.6\,\kmps$,
and a similar range is also found by \citet{mohanty:2005}. 
Fig.~\ref{fig:vrot_histrogram} shows the histogram
of $v\,\sin i$ distribution for the UScoOB objects (14 objects excluding
USco 121, non member). The log-normal fit of this distribution gives
the peak position at $16.9\,\kmps$ with a standard deviation $\sigma=27.8\,\kmps$.
Using the $v\,\sin i$ data in \citet{joergens:2001}, the same histogram
bin size used for UScoOB objects and a log-normal fit, we find the
$v\,\sin i$ distribution of the BD and VLM stars (8 objects) in Cha~I
peaks at $15.4\,\kmps$, and has the standard deviation of $8.0\,\kmps$.
The peak of the distribution is similar to that of Upper Sco objects,
but the standard deviation of the distribution is much smaller
than that of our Upper Sco objects. The difference maybe due to the
very small sample. A similar fit was applied to the $v\,\sin i$
distribution of 14 higher mass T~Tauri stars in Cha~I using the data
of \citet{joergens:2001}, and we found a peak at $17.0\,\kmps$ with a
standard deviation $25.9\,\kmps$ which are very similar to those of
the Upper Sco brown dwarf candidates and VLM stars.

\section{Binary fraction}
\label{sec:Binary-fraction}

In Section~\ref{sub:Radial-velocities}, we found 4 out of 17
(excluding USco~121; non-member) objects show a statistically
significant RV variation, indicating that they are binary/multiple
candidates. In order to estimate the uncertainty in the
binary/multiple fraction from this relatively small sample, we will
follow the method used by \citet{basri:2006} who considered the binomial
distribution, $P_{B}\left ( x, n, p \right )$ of $x$ positive event
out of $n$ trials with the probability $p$ for a positive event in each
trial. In our case, $n=17$ (the number of sample) and $x=4$.  The peak
of $P_{B}\left ( 4, 17, p \right )$ curve suggests the binary fraction of
24~per~cent, as its should be ($4/17=0.24$).  The uncertainties was estimated by
plotting $P_{B}\left ( 4, 17, p \right )$, and finding the values
of $p$ at which $P_{B}$ reduced to $e^{-1}$ of the peak value.  In this
analysis, we find the binary/multiple fraction along with the
uncertainties of our sample to be $f=24^{+16}_{-13}$~per~cent.

Next, we investigate the range of semimajor axes of binaries
(equivalently the range of binary periods) to which our RV survey is
sensitive. For this purpose, we will consider the detection
probability for binaries or the RV variables, given the time
separations of two-epoch observations and the ranges of estimated
primary masses (c.f.~Tables~\ref{tab:literatures} and
\ref{tab:summary_results}).  The probability is calculated based
on the simulated RV observations of binaries whose orbit are randomly
selected from a model. A similar method was used by
\citet{maxted:2005} and \citet{basri:2006}.  The most important
factors in determining the detection probability are the size of
uncertainties in RV measurements (which we use the average
$\sigma_{\mathrm{RRV}}$ from our observation), and the time
separations of observations.  The smaller the errors in RV
measurements, the larger the probability for a given binary orbit and
a time separation of orbit. The larger the time separation of
observations, the larger the upper limit of the semimajor axis to
which an observation is sensitive.  The average time separation of the
two-epoch RV observations of our targets is $21.6$~d.
In the following, we will briefly discuss our model assumptions and
parameters which are essentially the same as those of
\citet{maxted:2005} but with some simplifications.  

There are six basic parameters in our Monte Carlo simulation: 
primary mass ($M_{1}$), mass ratio ($q$), eccentricity
($e$), orbital phase ($\phi$), orbital inclination ($i$),
longitude of periastron ($\lambda$). The primary mass $M_{1}$ is
assigned from the adopted mass of 
the targets in Table~\ref{tab:literatures}, with a uniform random
deviation of $\pm0.002\,M_{\sun}$. The mass ratio is assumed to be
uniformly distributed between $q=1.0$ and $0.2$. The eccentricity $e$ is
assumed to be zero (circular orbits).  Both \citet{maxted:2005}
and \citet{basri:2006} found the detection probability is insensitive
to the assumed distribution of $q$ and $e$. The orbital phase is
randomly chosen between 0 and 1.  The inclination $i$ is randomly
chosen from the cumulative distribution of $\cos{i}$. The longitude of
periastron $\lambda$ is not necessary since we assumed $e=0$. 

In order to compute the detection probability as a function of
semimajor axis $a$, we take the following procedure: (1)~for each
object in our targets, we randomly select $10^5$ binaries using the
assumption stated above for a given value of $a$, and compute the RV
of the primary ($V_{1}$), (2)~evolve the orbit by the time separation
of the RV measurements used in the observations for this object, and
take another simulated measurement of RV ($V_{2}$), (3)~from $V_{1}$,
$V_{2}$ and the average uncertainty in RV ($\sigma_{\mathrm{RRV}}$)
from the observations, we compute the $\chi^{2}$ probability $p$,  and
flag the trial as a detection if $p<10^{-3}$ as done for the real
data, (4)~find the fraction of detections out of all random trials,
(5) repeat 1--4 for the range of $a$ between $10^{-3}$~and 10~au, and
(6) repeat 5 for the all targets, and find the detection probability
averaged over all targets as a function of $a$.

\begin{figure}

\begin{center}
\includegraphics[%
  clip,
  scale=0.63]{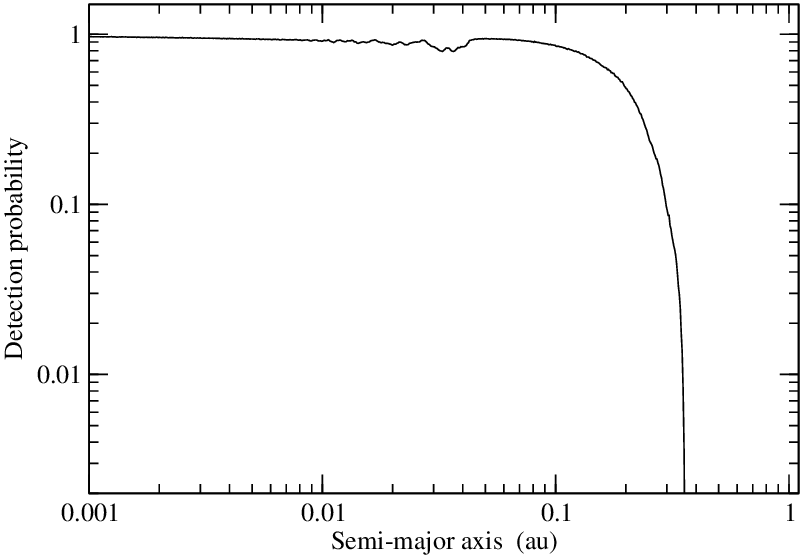}
\end{center}

\caption{The binary detection probability (solid) as a function of the semimajor
  axis ($a$), based on Monte~Carlo simulations of RV measurements, is
  shown.  For each object in our target list, the simulation was performed
  using the time separation ($\Delta t$) actually used in our
  two-epoch observations (c.f.~Table~\ref{tab:summary_results}) and
  the estimated primary mass  (c.f.~Table~\ref{tab:literatures}).  The
  final detection probability is obtained by averaging
  over the simulated observations of all the objects.  The average
  time separation in the two-epoch observations  is $\sim
  21.6\,\mathrm{d}$, and the average uncertainty in RV is
  $~0.42\,\kmps$. The probability remains fairly
  constant up to $a \approx 0.1\,\mathrm{au}$, and it rapidly decreases
  beyond $a\approx 0.3\,\mathrm{au}$.  The 80~per~cent detection
  probability is achieved for binaries with $a<0.1\,\mathrm{au}$. }  

\label{fig:detect_prob}

\end{figure}

The result of the simulation is shown in
Fig.~\ref{fig:detect_prob}. The detection probability curve shown here
is very similar that of \citet{basri:2006} for a constant time
separation (20~d) case (see their Fig.~4). The probability remains
fairly constant up to  $a \approx 0.1\,\mathrm{au}$, and it rapidly
decreases beyond $a\approx 0.3\,\mathrm{au}$.  This turning point will
increase if we had used larger time separations in our observation. 
From this figure, we find that the 80~per~cent detection probability
up to for binaries with $a<0.1\,\mathrm{au}$, given our time
separations and the average uncertainty ($\sim0.42\,\kmps$) in the RV
measurements. 

An alternative explanation for the RV variations found in the binary
candidates above is a type of stellar pulsations found by
\citet{palla:2005}. They studied the non-adiabatic, linear
instability of very VLM stars and BD during the deuterium burning phase
in the core, and found unstable fundamental modes in the time-scale
between $\sim 1$ and $\sim 5$~h for the object mass from $0.02$ and
$0.1\,\Msun$.  By using their pulsation periods and by assuming that the
pulsation amplitudes is 5~per~cent of the radius of the objects, we
find that the expected RV variation can be $>6\,\kmps$.
As mentioned in Section~\ref{sec:observation-reduction}, for each object at a
given night, two separate spectra are obtained consecutively (with
time separations of $\sim 0.5$~h.). We have measured the RV from each
spectrum to check if there is a large jump in the RV measurements on a
short time-scale. The size of the shifts in the RV values measured in
the two consecutive spectra are found in between $0.02\pm0.10$
(USco~128) and  $2.04\pm1.67\,\kmps$ (USco~66), and the average size
of the shifts $0.36\,\kmps$, which is much smaller than the 
expected RV variation based on the models of \citet{palla:2005};
hence, we found little indication of the pulsations in our sample.

\section{Conclusions}

\label{sec:Conclusions}

We have presented two-epoch RV survey of 18 young BDs and VLM stars
($0.02~M_{\sun} < M_{*} < 0.1~M_{\sun}$) in
UScoOB and $\rho$~Oph dark cloud core using the high resolution UVES
echelle spectroscopy at VLT.  The average time separation of RV
measurements are 21.6~d, and our RV survey is sensitive to binaries
with separation smaller than 0.1~au.  One of our targets, USco~121, is
most likely a non-member of the UScoOB association based on the
deviation of the RV from the result of the population in the
association. A similar conclusion was found by \citet{muzerolle:2003}
from their RV study and the low lithium abundance.  

We found 4 (USco~112, USco~128, USco~40 and USco~55) out of 17 objects as our 
binary/multiple candidates. This corresponds to the binary fraction of
$24^{+16}_{-13}$~per~cent for the binary separation $a<0.1$~au.  
The recent high-resolution imaging survey of brown dwarfs and very
low-mass objects (M5.5--M7.5 ) in the UScoOB by \citet{kraus:2005} (which was
not known to authors at the time of our observation: April--May, 2004)
confirms that USco~55 and USco~66 are multiple systems, and USco~109
is most likely a multiple system.  Their projected separations are
$a>4.0$~au which is well above our sensitivity limit of
0.1~au; therefore, it is not surprising that we did not identify USco~66 and
USco~109 as multiples. Interestingly, we also found USco~55 as 
a candidate for a multiple system. This may indicate that it is  
a triple system with one of the objects located within 0.1~au of the
primary.  In addition, we identify USco~112 and 128 as binary
candidates, but they did not find them as multiples. Further they found USco~67, 75,
130 and 132 as non-multiple, as did we.

We found the RV dispersion ($1.0\,\kmps$) of the objects in UScoOB is
very similar to that of the BDs and VLM stars in Chamaeleon~I (Cha~I)
previous study by \citet{joergens:2006b}. The rotational velocities
($v\,\sin i$) of the samples were also measured. The distribution of
$v\,\sin i$ for the UScoOB objects peaks around $16.9\,\kmps$ which is
also similar to that of the Cha~I population found by
\citet{joergens:2006b}; however, the dispersion of $v\,\sin i$ for the
UScoOB objects ($27.8\,\kmps$) is found to be much larger than that of
the Cha~I objects ($8.0\,\kmps$)

Follow-up spectroscopic observations of the binary candidates
presented here are planned in near future. There are only a few RV
variable binary candidates identified in earlier surveys
(\citealt{guenther:2003}; \citealt{kenyon:2005}; \citealt{joergens:2006}).
As \citet{burgasser:2006} points out most of the current RV and
imaging surveys use samples from magnitude-limited survey, but one
should attempt to use the samples from volume-limited survey in order
to find a correct statistics on binary parameters more straightforwardly,
i.e. without correcting for bias. 
\textcolor{black}{
It is possible that our binary fraction is an overestimate since we
are biased to brighter objects. However the lack of a double-line
system suggests that a primary-to-secondary luminosity ratio of $\gg 1$
for the RV variables found in this paper. 
}

\section*{Acknowledgements}

We would like to thank the anonymous referee for constructive
comments and suggestions for improving the clarity of the manuscript.
We also thank the staff of VLT of the ESO for carrying out the observations
in service mode. RK is grateful for Rob Jeffries  and Tim Naylor for helpful
suggestions on the data analysis presented in this paper. We also
thank Matthew Bate for providing us valuable comments on
the manuscript. This work was supported by PPARC rolling grant
PP/C501609/1.

%
%


\label{lastpage}
\end{document}